# IMPACT OF NUMBER OF ELEMENTS ON THE DIRECTIVITY OF PLANAR ARRAY OF MONOPOLE ANTENNA

**Akpo Simon (PhD)** [International University of Management, Namibia | s.akpo@ium.edu.na]
**Omini Ofem** [University of Cross River State, Nigeria | ofemuket0@gmail.com  ]
**Tawo Godwin** [University of Cross River State, Nigeria | edwardtawomeji@gmail.com]

**Abstract :** This research investigates how the number of elements affects the monopole antenna's planar array's directivity. This study also takes into account the antenna's effect on the whole field it radiates. The monopole antennas are arranged in a planar configuration with all the components in their proper locations using the Hadamard matrix approach. Each matrix's directivities and array factors were calculated, and a MATLAB tool was used to simulate the radiation pattern. A range of elements from 4 X 4 to 50 X 50 planar layouts were taken into consideration during the investigation. Increasing the number of elements improves the directivity. Increasing the number of elements in the planar array resulted in a great improvement in directivity, as seen by the computed and simulated results. Consequently, by increasing the antenna's directivity, a greater number of elements influences the overall field emitted.

**Keywords:** Grating lobes, Array Antenna, Planar Array, and Array factor, Directivity and Element,

## 1. Introduction

In order to lessen the inherent grating lobes and lobe levels, multiple antennas are connected and placed in a matrix fashion on a plane, creating planar array antennas [1 2]. On the other side, directivity, a basic antenna property, quantifies how "directional" an antenna's radiation pattern is [1,4,5]. The communication sector has a strong need to transmit signals over great distances while maintaining very high signal quality and strength, and in recent years, numerous antenna synthesis methods have been developed for planar arrays. The creation of array antenna technology [1] was motivated by the inability of single antennas, such as monopoles, dipoles, and others, to offer the requisite signal quality at such a distance [6]. The monopole antenna with a planar array was selected for this study because of its capacity to send signals over vast distances while maintaining the signal's power and content quality. This planar array of monopole antenna's total field radiated and directivity are mostly determined by the elemental parameters (inter-element spacing, number of elements, and progressive phase of the array elements). However, the directivity of the radiated field can be improved and enhanced by a number of factors [1].

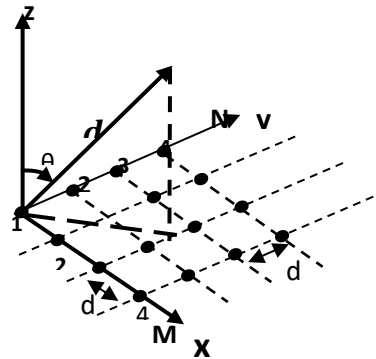

Fig. 1. The geometry of N × M planar Array Antenna [1]

The design principles for planar arrays antennas are similar to those elements placed in two dimensions Fig. 1, the array factor of a planar array is the multiplication of the array factors of two linear arrays, one along the x-axis and the other one along the y-axis. Therefore, planar array factor $AF_{planar}$ is simply expressed as:

$$AF_{planar} = \left\{\frac{1}{M}\frac{\sin\left(\frac{M}{2}\rho_x\right)}{\sin\left(\frac{\rho_x}{2}\right)}\right\}\left\{\frac{1}{N}\frac{\sin\left(\frac{N}{2}\rho_y\right)}{\sin\left(\frac{\rho_y}{2}\right)}\right\} \quad (1)$$

Where,

$$\rho_x = ka_x\sin\theta\cos\phi + \omega_x \quad (2)$$

$$\rho_y = ka_y\sin\theta\cos\phi + \omega_y \quad (3)$$

Where $\rho_x$ and $\rho_y$ are the array phase functions along $x$ and $y$ axis, respectively. $\omega_x$ and $\omega_y$ is the progressive phase shift along $x$ and $y$ axis respectively. $a_x$ and $a_y$, is the inter-element spacing along $x$ and $y$ axis respectively. Given that the phase factor $\rho_x$ and $\rho_y = 0$, then we have:

$$\omega_x = -ka_x\sin\theta\cos\phi \quad (4)$$

$$\omega_y = -ka_y\sin\theta\cos\phi \quad (5)$$

And

$$k = 2\frac{\pi}{\lambda} \quad (6)$$

Equation (6), is the free space wave constant. The nulls of the array function are found by determining the zeros of the numerator term where the denominator is not simultaneously zero [2 10]. That is;

$$\sin\left(\frac{N\rho}{2}\right) = 0 \quad \Rightarrow \quad \frac{N\rho}{2} = \pm n\pi \quad \Rightarrow \quad \phi + ka\cos\theta_n = \pm\frac{2n\pi}{N} \quad (7)$$

$$\theta_n = \cos^{-1}\left[\frac{\lambda}{2\pi a}\left(-\alpha \pm \frac{2n\pi}{N}\right)\right] \quad (8)$$

$$n = 1,2,3,\ldots$$

$$n \neq 0, N, 2N, 3N, \ldots$$

The peaks of the array function can be found by determining the zeros of the numerator term where the denominator is simultaneously zero.

$$\theta_m = \cos^{-1}\left[\frac{\lambda}{2\pi a}(-\alpha \pm 2m\pi)\right] \quad m = 1,2,3\ldots \quad (9)$$

When $m = 0$ term,

$$\theta_m = \cos^{-1}\left(\frac{\lambda}{2\pi a}\right) \quad (10)$$

For a beam pattern broadside

$d_x = d_y = \frac{\lambda}{2}$, $\omega_x = \omega_y = 0$ and $M = N = 4$

Let $\rho_x = kd\cos\theta + \omega = \rho_y$ for a uniform array and $M = N = 4$. $\omega_x = \omega_y = 0, \theta = 45°$

### 2. The Wavelength (Lambda, $\lambda$)

It has been established that the speed of electromagnetic wave on air is $3 \times 10^8 m/s$, and can be obtained [12] from the formula,

$$V = f\lambda \quad (11)$$

Where,

f = the frequency of the electromagnetic wave

$\lambda$ = Lambda, the wavelength

For the purpose of this investigation, the frequency is 2.5GHz, therefore, the wavelength, $\lambda$ can be obtained as follows;

$$\lambda = \frac{v}{f} = \frac{3 \times 10^8}{2.5 \times 10^9} = \frac{3 \times 10^{-1}}{2.5} = 0.12m$$

The (AF) in this research is calculated using the normalized (AF), [2] of equation (12) as;

$$AF_n(\theta,\phi) = \left\{\frac{1}{M}\frac{\sin\left(\frac{M}{2}\rho_x\right)}{\sin\left(\frac{\rho_x}{2}\right)}\right\}\left\{\frac{1}{N}\frac{\sin\left(\frac{N}{2}\rho_y\right)}{\sin\left(\frac{\rho_y}{2}\right)}\right\} \quad (12)$$

Where,

$$\rho_x = ka_x\sin\theta\cos\emptyset + \omega_x$$

$$\rho_y = ka_y\sin\theta\cos\emptyset + \omega_y$$

For a beam pattern broadside

$d_x = d_y = \frac{\lambda}{2}$, $\omega_x = \omega_y = 0$ and $M = N = 4$

Let $\rho_x = kd\cos\theta + \omega = \rho_y$ for a uniform array and $M = N = 4$. $\omega_x = \omega_y = 0, \theta = 45°$

## 3. Determination of the Directivity $D_x$ and $D_y$ of a Linear Array

The directivity of linear array antenna D, is determined by the formula,

$$D = \frac{2R_0^2}{1 + (R_0^2 - 1)f\frac{\lambda}{(L+a)}} \qquad (13)$$

Where, $R_0$ is the voltage ratio, $L$ is the length of the linear array, $f$ is the broadening factor of broad array and $a$ is the distance between elements in the linear array [2]. Let $R_0 = 20$ voltage ratio, $L = 4, a = 0.47, f$

## 4. Determination of the Directivity $D_0$ of the Array Antenna

Since the configuration of a planar array antenna consists of two linear antennas placed orthogonally to each other, the directivity of a planar array antenna, $D_0$ can be obtained as [2];

$$D_0 = D_x D_y \cos\theta \qquad (14)$$

Where,

$D_x$ = The directivity of a linear array along x – axis

$D_y$ = The directivity of a linear array along y – axis

Table 1 is constructed by the computation of varying the number of elements of the Planar Array of Monopole Antenna in both $x$ and $y$-directions keeping progressive phase shift, theta/phi = 45 and inter-elements spacing = $0.47\lambda$ constant throughout the variation. The table contain the values of number of elements along both $x$ and $y$-directions

**Table 1 Result of Varying the number of elements on the array factor**

| S/N | No. of Elements Array | $AF_n(\theta)_X$ | $AF_n(\theta,\emptyset)_{xy}$ | $AF_n(\theta)_{dB}$ |
|---|---|---|---|---|
| 1 | 4 | 0.9992 | 0.9984 | -0.0072 |
| 2 | 5 | 0.9980 | 0.9980 | -0.0176 |
| 3 | 6 | 0.9981 | 0.9978 | -0.0168 |
| 4 | 8 | 0.9834 | 0.9674 | -0.1452 |
| 5 | 10 | 0.9944 | 0.9888 | -0.0489 |
| 6 | 15 | 0.9876 | 0.9754 | -0.1079 |
| 7 | 20 | 0.9781 | 0.9567 | -0.1927 |
| 8 | 25 | 0.9658 | 0.9323 | -0.3021 |
| 9 | 30 | 0.9510 | 0.9044 | -0.4379 |
| 10 | 35 | 0.9336 | 0.8716 | -0.5966 |
| 11 | 40 | 0.9138 | 0.8350 | -0.7827 |
| 12 | 45 | 0.8917 | 0.7951 | -0.9957 |
| 13 | 50 | 0.8673 | 0.7522 | -1.2363 |

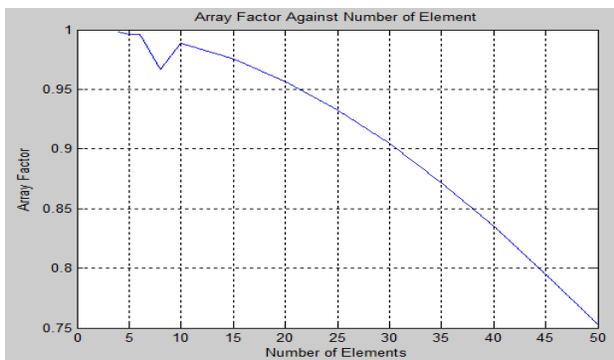

(a)

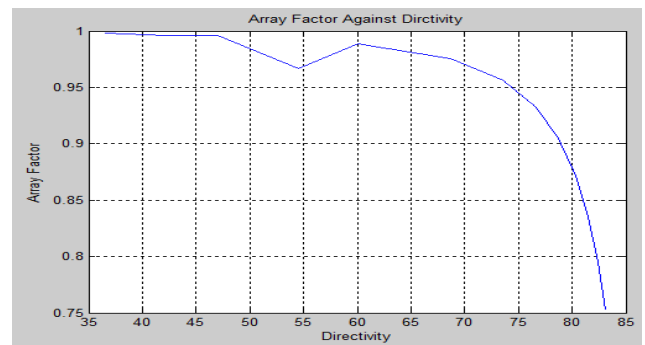

(b)

Fig. 2. Graph of Array Factor Against a) Number of Element and b) Directivity

The above Fig. 2a and 2b are obtained from plotting Array Factor against Directivity, and Number of elements of the Table 1 above. The two graphs obtained shows that as the Number of elements of the matrix array is increased, the Array Factor is decreases as well as the Directivity.

## 5. Radiation Patterns of Varying Number of Elements

This section shows the various radiation patterns obtained by varying the number of elements in the matrix array. While the values angles theta and phi is 45° and the inter-element spacing for both $x\ and\ y\ axis$ is $0.47\lambda$.

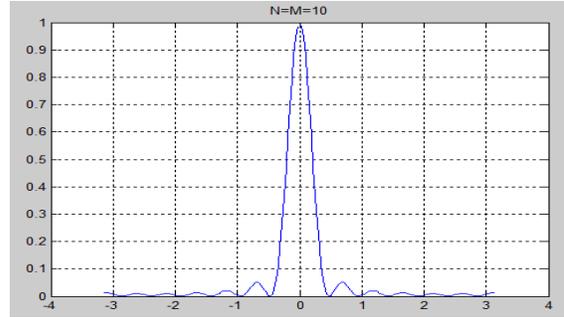
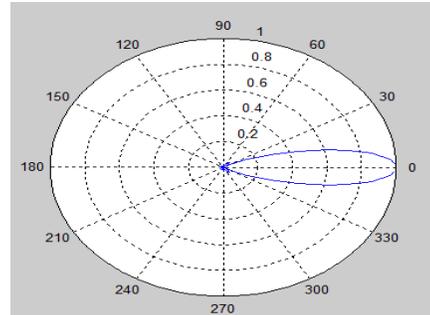

$10\ X\ 10$

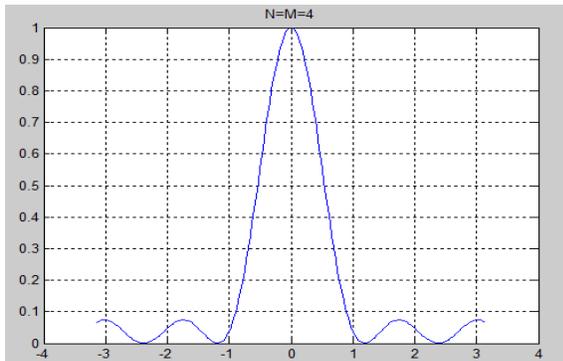
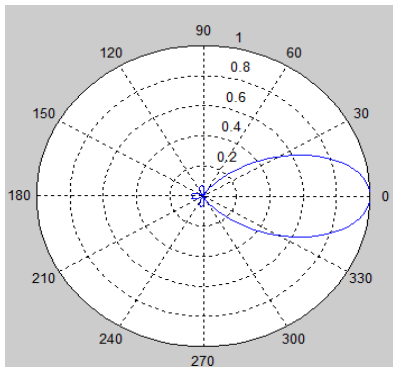

$4\ X\ 4$

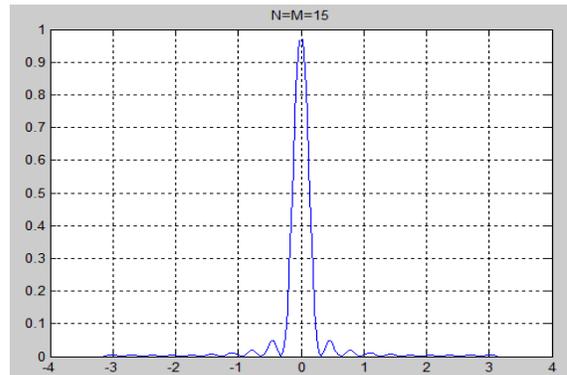
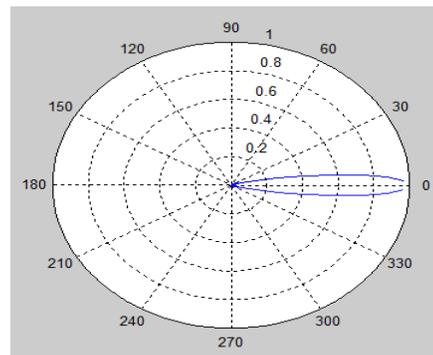

$15\ X\ 15$

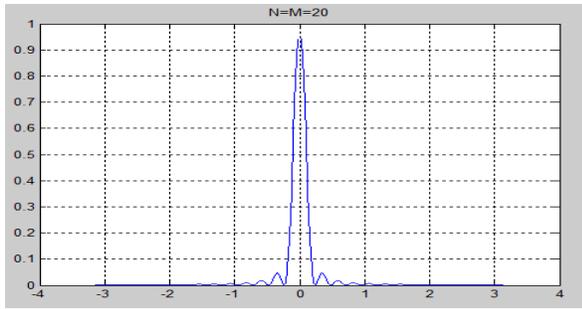
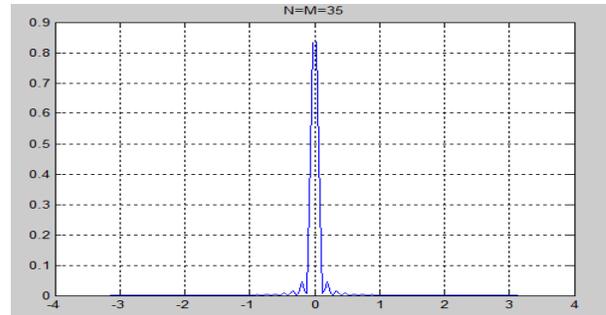
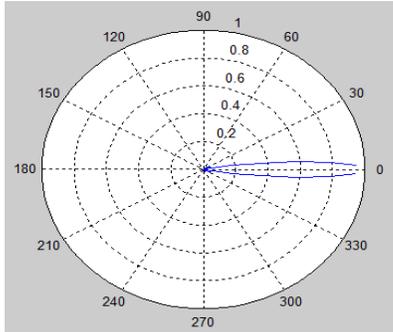
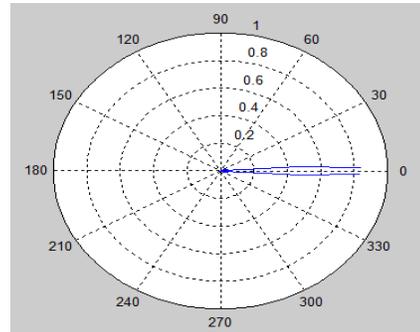

20 $X$ 20

35 $X$ 35

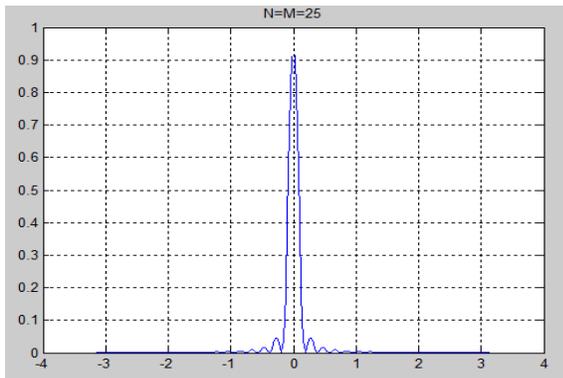
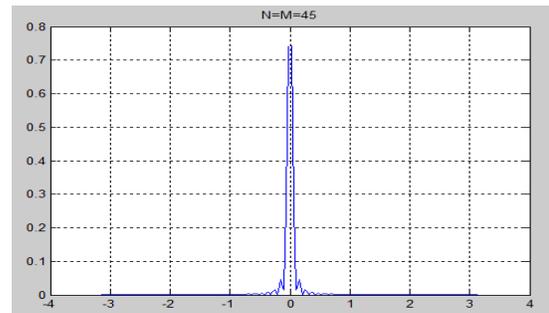
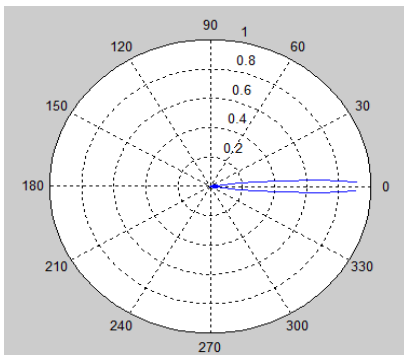
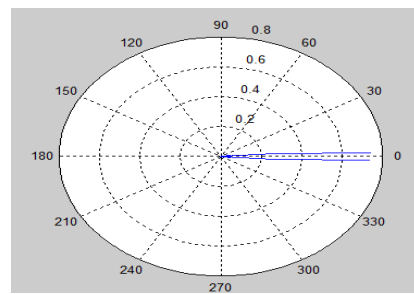

45 X 45

25 $X$ 25

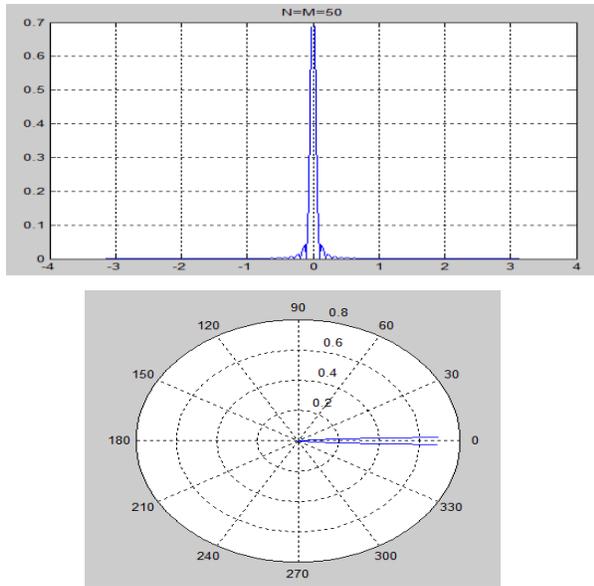

50 X 50

**Fig. 3 Results of Varying the number of elements on the Array Factor and Directivity**

## 6. Discussion of Results

According to Table 1, as the number of elements increases, the directivity continuously decreases. For example, the array factor for a matrix with four elements is equal to the list elements' value, or 0.9984. However, when fifty elements were taken into account, 0.7522 was found. Conversely, when the number of elements rises, the directivity does the opposite.

13 planar array antennas with N X N matrices of 4 X 4, 6 X 6, 8 X 8, 10 X 10, 15 X 15, 20 X 20, 25 X 25, 30 X 30, 35 X 35, 40 X 40, 45 X 45, and 50 X 50 are shown in Fig. 3 for their directivity of radiation pattern. For every array, the steering phase angle is 45° for the theta and phi constants, and the element spacing is 0.47λ for the x and y axes. It is evident that as the array's elements rise, so does its directivity. Beside the primary lobes, there is no growth of side lobes. Additionally, as the matrix size grows and the primary lobe gets more directed, its size decreases. The 50 X 50 elements matrix yields the best directivity in this paper.

## 7. Conclusion

This work examines the effects of the number of elements on the array factor and the directivity of the planar array of monopole antenna. The goal was to use a monopole antenna to examine how changing the number of elements affected the antenna's directivity. Directivity of the radiated field increases as matrix size grows, as seen by Table 1 and Figure 1 results. Side lobe development is absent. Additionally, the major lobe gets more directed and its size decreases as the array's elements grow from 4 X 4 to 50 X 50. As a result, increasing the planar array's matrix size influences the array factor, the antenna's overall field of radiation, and the main lobe's directivity in the radiation pattern.

**Acknowledgements**: Akpo (PhD) is the head of research and senior lecturer at the International University of Management (IUM), Namibia. Tawo, and Omini, are Electrical and Electronic Engineering lecturers at the University of Cross River State, Nigeria.

**Funding:** The authors declare that there was no funding given in support if this article.

**Authors' Contributions:** OFU is the lead author, and TGA performed the experiment and data analysis. SEA is the content editor and chief correspondence.

**Competing interests:** The authors declare that there were no competing interests.

**Availability of data and materials:** The datasets generated and/or analysed during the current study are available on request.